\documentclass{rQUF2e}
\graphicspath{{Figures/}}

\newlength{\figurewidth} 
\newlength{\figureheight} 

\begin{document}

\title{Efficient computation of mean reverting portfolios using cyclical coordinate descent}

\author{T. GRIVEAU-BILLION $^{\ast}$$\dag$\thanks{$^\ast$Corresponding author.
Email: theophile.griveau-billion14@imperial.ac.uk} and B. CALDERHEAD${\dag}$\\
\affil{$\dag$ Department of Mathematics, Imperial College London, 180 Queen's Gate, Kensington, London SW7 2AZ, UK\\} \received{v0 released April 2019} }

\maketitle

\begin{abstract}

The econometric challenge of finding sparse mean reverting portfolios based on a subset of a large number of assets is well known.
Many current state-of-the-art approaches fall into the field of co-integration theory, where the problem is phrased in terms of an eigenvector problem with sparsity constraint.  Although a number of approximate solutions have been proposed to solve this NP-hard problem, all are based on relatively simple models and are limited in their scalability.
In this paper we leverage information obtained from a heterogeneous simultaneous graphical dynamic linear model (H-SGDLM) and propose a novel formulation of the mean reversion problem, which is phrased in terms of a quasi-convex minimisation with a normalisation constraint.  This new formulation allows us to employ a cyclical coordinate descent algorithm for efficiently computing an exact sparse solution, even in a large universe of assets, while the use of H-SGDLM data allows us to easily control the required level of sparsity.  We demonstrate the flexibility, speed and scalability of the proposed approach on S\&P$500$, FX and ETF futures data.

\end{abstract}


\begin{classcode} C11, C32, G11 \end{classcode}

\clearpage
\section{Introduction}
Finding a linear combination of stocks that mean revert is an econometric problem that has been captivating researchers for years, since such portfolios with reasonably predictable short term behaviour can be used for developing systematic trading strategies.  When applied to two assets, this is often referred to as looking for trading pairs.  Such combinations however may be hard to find due to non-stationarity and combinatorial considerations, especially for baskets of more than two assets.  Furthermore, finding a mean reverting portfolio is not sufficient in itself; indeed, to be viable the effect must persist over a long enough time horizon and be strong enough to overcome trading costs.\bigskip

Currently, the classic approach to this problem relies on cointegration theory (see \cite{Johansen:2005wh} for an overview), which tries to find a linear combination of non-stationary processes that is stationary.  These methods usually rely on statistical tests to check that the time series are cointegrated.  However, as stated in \cite{Johansen:2005wh} such tests strongly depend on the modelling assumptions of the data.  In this paper we are interested in the case of choosing a sparse subset of more than two assets from a much larger universe of assets. Recent advances in this area have proposed a new formulation of the problem \citep{dAspremont:2010kh, Cuturi:2013uj, Cuturi:2015wg}, which relies on a characterisation of portfolio predictability \citep{Box:1977iz}.  With this definition, a mean reverting portfolio (MRP) is characterised by a low predictability, in contrast to a momentum portfolio which has high predictability. In these three papers, the authors reformulated the MRP definition as an optimisation problem with constraints and compared it to other methods, such as minimising the portmanteau test or minimising the crossing statistic.  They emphasised the importance of a good MRP to be sparse and argued in favour of an additional minimum variance constraint to select a portfolio with a variance high enough for the effect to be tradable.\bigskip

The approach of \cite{dAspremont:2010kh} and \cite{Cuturi:2013uj, Cuturi:2015wg} relies on the hypothesis of \cite{Box:1977iz}, which requires a model for the price process $s_t$ conditional on all previous information.  The authors assumed that the assets follow a vector autoregressive model, $VAR(p)$, of order $p$ and that the covariance matrix of the portfolio be stationary.  In \cite{dAspremont:2010kh}, and subsequently \cite{Sipos:2013ke} and \cite{Fogarasi:eh}, they also assumed that the resulting portfolio follows an Orstein-Uhlenbeck process and made the parallel between the mean reversion parameter and the predictability variable.  In order to solve this optimisation, \cite{Box:1977iz} rewrote the problem as a generalised eigenvalue problem, to which \cite{dAspremont:2010kh, Cuturi:2013uj, Cuturi:2015wg} then added a sparsity constraint, making this combinatorial problem NP-hard, and proposed to solve this minimisation with a semi-definite relaxation approach.  As an alternative, \cite{Sipos:2013ke, Fogarasi:eh} proposed a simulated annealing algorithm to find the solution.  Both methods produce approximations of the solution, are expensive to compute and lack scalability.  Finally, given the resulting MRP, \cite{Cuturi:2013uj, Cuturi:2015wg} followed a strategy introduced by \cite{Jurek:2007ju} to trade a stationary autoregressive processes, whereas \cite{Sipos:2013ke, Fogarasi:eh} assumed the portfolio follows an Ornstein-Uhlenbeck process and thus used decision theory to implement their trading strategy.\bigskip

In this paper we extend the approach of \cite{dAspremont:2010kh, Cuturi:2013uj, Cuturi:2015wg} by characterising a sparse mean reverting portfolio according to its predictability but without assuming a VAR$(p)$ process.  Instead we will leverage the results of a heterogeneous simultaneous graphical dynamic linear model (H-SGDLM) to set up the MRP optimisation problem for assets following time varying autoregressive TVAR$(p)$ processes.  The H-SGDLM builds a sparse covariance matrix which is not assumed stationary and our new formulation of the MRP minimisation problem with normalisation constraint allows us to use an efficient optimisation to compute the global minimum.\bigskip

The H-SGDLM of \cite{GriveauBillion:2019gf, Gruber:2016wn, Corsi:2004eg} models the entire investment environment by decomposing each time series, and thus the whole market, into economically-meaningful variables.  This decomposition creates different groups of variables representing the endogenous and exogenous influences on the behaviour of each asset.  Many different signals are created by this model; for this paper we are mainly interested in the graph of connections between the different assets and the resulting sparse covariance matrix and autoregressive coefficients.  These variables will allow us to propose a new formulation of the sparse mean reverting portfolio optimisation problem.\bigskip

While the previous approaches of \cite{dAspremont:2010kh, Cuturi:2013uj, Cuturi:2015wg} and \cite{Sipos:2013ke, Fogarasi:eh} obtained approximations of the solution to a constrained optimisation problem, we propose a new algorithm that finds the global minimum.  We employ a cyclical coordinate descent (CCD) algorithm which is fast and efficient in large-scale problems.
Furthermore, \cite{Tseng:2001bd} proved the convergence to the global minimum of cyclical coordinate descent algorithms when the the utility function can be written as the combination of a convex differentiable function and a non-differentiable but additively separable one.\bigskip

Building on this result, \cite{Anonymous:2014di} used the CCD algorithm to solve the mean variance portfolio with an elastic-net constraint. In addition, \cite{GriveauBillion:2013to} used this CCD method to solve the non-convex equal risk contribution portfolio optimisation problem, where it was observed to be faster than the Newton algorithm in large environments. We follow the direction of these approaches and derive a CCD algorithm to solve the sparse MRP minimisation problem with a normalisation constraint. The algorithm we propose can be applied to compute a MRP from any predicted covariance matrix. However, we focus in this paper on the predictions obtained from the H-SGDLM model.\bigskip

Using this new MRP solution we can then consider a number of different trading strategies. Hence we will numerically compare the approaches proposed by \cite{Jurek:2007ju} and \cite{Sipos:2013ke, Fogarasi:eh}, as well as a combination of the two. In order to assess the performance and robustness of our solution we computed the sparse MRP and implemented the trading strategy on three different datasets: stocks from the S\&P$500$, futures on FXs and futures on ETFs. All of the backtests we obtained exhibit a linear growth throughout the time under study, which includes the financial crisis of 2008.  While in some environments the backtests are smoother than others, the fact that none has a particularly large draw-down, and they are all roughly linear through time, highlights the robustness of our proposed approach.\bigskip

Section \ref{sec:MRP_d} will recall the sparse mean reversion optimisation problem and briefly review the current approaches for solving it.  Section \ref{sec:newMRP} will then detail the H-SGDLM model and the novel formulation of the MRP problem it allows.  The CCD algorithm and its application to solve the new MRP optimisation are detailed in section \ref{sec:CCD}. Once the optimal solution is found we consider possible trading strategies, as detailed in section \ref{sec:trading}. Finally, in Section \ref{sec:results} we present results for different assets classes.

\section{Description of the sparse MRP optimisation problem}
\label{sec:MRP_d}
\subsection{The MRP as a constraint minimisation problem}
In this paper we build on the approach introduced by \cite{dAspremont:2010kh, Cuturi:2013uj, Cuturi:2015wg} for characterising a mean reverting portfolio. The authors used the predictability ratio defined for stationary processes by \cite{Box:1977iz} and extended it by adding a cardinality constraint and a minimum variance constraint, in \cite{dAspremont:2010kh} and \cite{Cuturi:2013uj, Cuturi:2015wg}, respectively. Let us assume a univariate model such that the asset $s_t$ follows the process $s_t = \hat{s}_t + \epsilon_t$, where $\epsilon_t \sim \mathcal{N}(0, \sigma_{\epsilon}^2)$ represents i.i.d. noise, and $\hat{s}_t$ is an estimator for $x_t$, which can use all the information available up to $t-1$. With this formulation, \cite{Box:1977iz} used the variance $\sigma^2 = \hat{\sigma}^2 + \sigma_{\epsilon}^2$ to define the predictability as the ratio of the variance of the estimator over the variance of the asset:
\begin{equation*}
\nu = \frac{\hat{\sigma}^2}{\sigma^2} \, .
\end{equation*}
The interpretation motivating the use of this ratio is as follows: if the variance of the estimator $\hat{\sigma}^2$ is higher than that of the asset $\sigma^2$, then the dependence on the past dominates the noise and thus the predictability is high, while if $\nu$ is small then the noise dominates the prediction. It is straightforward to extend this definition to a multivariate model where $\nu$ becomes the ratio of the variances of two portfolios. The \cite{Box:1977iz} procedure requires us to define a model for the price process; in \cite{dAspremont:2010kh, Cuturi:2013uj, Cuturi:2015wg} the authors considered the prices to follow a VAR$(1)$ process: 
\begin{equation*}
S_t = A S_{t-1} + \epsilon_t \, .
\end{equation*}
where $S_t$ is a vector of prices, $A$ the autoregressive coefficient and $\epsilon_t \sim \mathcal{N}(0, \Sigma)$ the noise with covariance matrix $\Sigma$. For the multivariate case, we use a weight vector $x_t$ to form the portfolio $x_t^T S_t$ and obtain the predictability ratio:
\begin{equation*}
\nu = \frac{x_t^T A^T \Sigma A x_t}{x_t^T \Sigma x_t} \, .
\end{equation*}
\cite{Box:1977iz} showed that minimising the predictability is equivalent to solving the generalised eigenvalue problem, $\text{det}(\nu \Sigma - A^T \Sigma A) = 0$.  \cite{dAspremont:2010kh, Cuturi:2013uj, Cuturi:2015wg} highlighted the fact that we only need to find the extremal generalised eigenvalues to rewrite the problem as a minimisation. They used this new formulation as an optimisation problem to introduce a sparse constraint on the weight vector, however the resulting combinatorial problem has been proven to be a NP-hard. We can write this minimisation problem with normalisation and sparse constraint as:
\begin{eqnarray}
\label{eq:minProb}
\text{min}_{x_t} \text{\;\;} \frac{x_t^T A^T \Sigma A x_t}{x_t^T \Sigma x_t} \text{\;\;\; subject to: } && \| x \|_2 = 1  \, , \\
&& \| x \|_0 = k  \, .
\end{eqnarray}
In \cite{Cuturi:2013uj, Cuturi:2015wg} the authors reformulated this optimisation to include a constraint on the minimum variance of the portfolio and proposed the use of two other utility functions, which have fewer constraining assumptions: the Portmanteau statistic, which tests if a process is white noise, and the crossing statistic, which represents the frequency at which a process crosses its mean.  In this paper however we will focus solely on the predictability ratio, since its formulation is suited for the H-SGDLM model and the CCD algorithm. \bigskip

\cite{Sipos:2013ke, Fogarasi:eh} followed the idea of \cite{dAspremont:2010kh} but considered the inverse problem. Indeed, they assumed the portfolio follows an Ornstein-Uhlenbeck process and considered the coefficient representing the mean reversion speed of the portfolio to be directly proportional to the predictability variable, as defined by \cite{Box:1977iz}. Hence they maximised the predictability instead of minimising it. In a second paper \cite{Sipos:2013ke} proposed replacing the generalised eigenvalue maximisation problem introduced by \cite{dAspremont:2010kh} with a maximisation of the average return.  This seems to highlight the fact that they compute the inverse of a mean reverting portfolio and, as explained by \cite{dAspremont:2010kh}, by maximising the predictability they obtained a momentum portfolio.  For this reason we follow the approach of \cite{Box:1977iz} and \cite{dAspremont:2010kh, Cuturi:2013uj, Cuturi:2015wg} in the rest of this paper.\bigskip

Since \cite{dAspremont:2010kh, Cuturi:2013uj, Cuturi:2015wg} modelled the price process as a VAR$(1)$ process, the authors had to estimate $A$, the autoregressive coefficient, and $\Sigma$, the covariance matrix, in order to use the \cite{Box:1977iz} formulation of predictability.  \cite{Box:1977iz} considered the least squares estimate for $A$ and plugged this into their general eigenvalue problem. On the other hand, \cite{dAspremont:2010kh, Cuturi:2013uj, Cuturi:2015wg} added an $L_1$ penalty to obtain more stable estimates and highlight the dependencies between $S_t$ and $S_{t-1}$. Hence, they used a likelihood with $L_1$ penalty to compute the covariance matrix and replaced the least squares estimate of $A$ with a LASSO-based estimate. In \cite{Cuturi:2013uj, Cuturi:2015wg} the $lag-k$ covariance matrices were computed according to their empirical estimates and the minimisation problem was reformulated with those variables. 

\subsection{Previous techniques used to solve this optimisation}
Since the MRP optimisation problem is non-convex and includes a cardinality constraint, the classic minimisation algorithms will not work. In their papers, \cite{dAspremont:2010kh, Cuturi:2013uj, Cuturi:2015wg} solve the constraint optimisation problem using semi-definite programming.  In particular they relax the problem into a convex semi-definite optimisation programme (SDP), which they then solve with a minimum eigenvalue solver.  However, this relaxation only provides a sub-optimal solution.  Indeed, while they proved that in special cases the relaxed problem is tight, i.e. gives the exact solution, in general the equivalence is not guaranteed.  Nonetheless, when the correspondence is not exact, the solution of this semi definite relaxation can be deflated to produce a good approximation.  In order to do so, the authors used a sparse PCA method which takes the SDP solution and converts it into a sparse vector of weights by recovering the leading sparse eigenvectors.\bigskip

While we do not fully agree on the utility function proposed by \cite{Sipos:2013ke, Fogarasi:eh}, the methods the authors use to solve the resulting optimisation problem are nonetheless interesting. Indeed, they proposed two techniques: a simulated annealing (SA) approach and a feed forward neural network (FFNN) algorithm.  In their paper the authors also compared them with exhaustive search methods on small enough problems, before using them to tackle bigger ones where those greedy approaches are too expensive to run. For the SA method they converted the problem into a combinatorial one by considering only integer values for the weights $x$. Thus, they run the SA algorithm on a discrete grid whose dimension corresponds to the number of assets. As a stochastic optimisation procedure this is not guaranteed to find the optimal solution, however, and its performance depends heavily on the choice of the initial vector.  For the FFNN they first considered the problem in a smaller dimension and computed some optimal portfolio via exhaustive search to create a training dataset. Once the network has been trained with classic back-propagation on this sub-problem they run it on the large one to obtain a weight vector, again with no guarantees of convergence. The results of the FFNN algorithm were also used as an initialisation procedure to select the weight vector for the simulated annealing steps.\bigskip

These different approaches incorporate the cardinality constraint at different steps of the algorithm to obtain the closest approximation to the optimal solution of this non-convex problem.  Unfortunately all of these methods provide only an approximation of the optimal weight vector and are expensive to compute.  Furthermore, only the SDP technique can prove an optimal convergence in some special cases. In the coming section we will introduce a novel formulation of the MRP optimisation problem and a procedure to compute the weight vector which is guaranteed to converge and works efficiently even in a large environment.

\section{New formulation of the MRP with H-SGDLM}
\label{sec:newMRP}
\subsection{Description of the H-SGDLM model}
\label{sec:HSGDLM}
The Bayesian multivariate modelling of the market, H-SGDLM model of \cite{GriveauBillion:2019gf, Gruber:2016wn, Corsi:2004eg} gives interesting insights on the behaviour of each asset and the rest of the market. The letters of H-SGDLM stands for heterogeneous simultaneous graphical dynamic linear model. This model extended the previously introduced SGDLM model of \cite{Gruber:2016wn} and HAR-RV model of \cite{Corsi:2004eg} to decompose the moves of the time series into different economically-meaningful variables which can be grouped into endogenous and exogenous ones. In addition, it does not assume stationarity of the time series and sequentially updates its parameters with each newly available data point thus producing new signals on the evolution of the different factors. Furthermore, the flexibility of this model makes it straightforward to extend it to fit our particular problem.\bigskip

The H-SGDLM models each time series independently and then combines them to form a complete multivariate model of the market hence its matrix formulation:
\begin{eqnarray*}
S_t &\sim& \mathcal{N}\left( H_t \mu_t, \Sigma_t \right) \, , \\
\text{With:} && H_t = \left( I - \Gamma_t \right)^{-1} \, , \\
&& \Omega_t = \Sigma^{-1} = \left( I - \Gamma_t \right)^T \Lambda_t  \left( I - \Gamma_t \right) \, .
\end{eqnarray*}
Where $\Lambda_t$ is the diagonal precision matrix with $\lambda_{j,t}$ on its diagonal, $\Gamma_t$ is the matrix of the exogenous state coefficients, and $\mu_{j,t} = e_{j,t} \phi_{j,t}$ the mean vector. Thus, $\Omega_t$ corresponds to the sparse precision matrix and $\Sigma_t$ the corresponding sparse covariance matrix.\bigskip

The endogenous group of variables represents the influence of the previous information of this time series while the exogenous one represents the influence of the rest of the market. Hence, the endogenous variables include, among others, the leverage effect and the influence of lower frequencies on the current evolution. On the other hand, the exogenous variables combine an evolving set of assets from the market which have been selected for their influence on the time series. While in the paper \cite{GriveauBillion:2019gf} they mainly used the H-SGDLM to model and predict the variance of individual time series and the market, in this paper we will apply it directly on the logarithm of stock prices to obtain the variables needed to solve the mean reversion portfolio optimisation.\bigskip

For asset $j$ the time series of prices $s_{j,t}$ is modelled as a function of the state variables; both assuming a Normal Inverse-Gamma model. Let us denote by $\theta_{j,t}$ the state vector, $F_{j,t}$ the vector of variables and $G_{j,t}$ the state evolution matrix; the model reads:
\begin{eqnarray*}
s_{j,t} &=& F_{jt} \theta_{j,t} + v_{j,t} \, , \\
\theta_{j,t} &=& G_{j,t} \theta_{j,t - 1} + \omega_{j,t} \, , \\
\text{with:} && v_{j,t} \sim \mathcal{N} (0,\lambda_{j,t}^{-1}) \, , \\
&& \omega_{j,t} \sim \mathcal{N}(0,W_{j,t}) \, .
\end{eqnarray*}
Where the variance of the states $W_{j,t}$ and of the observations $\lambda_{j,t}^{-1}$, are both assumed to follow an Inverse-Gamma distribution. The vectors of variables $F_{j,t}=\left(e_{j,t}, r_{sp_t(j), t} \right)$ and states $\theta_{j,t} = \left( \phi_{j,t}, \gamma_{j,t} \right)$ are composed of two subsets representing the endogenous and exogenous variables and coefficients, respectively. The endogenous variables represent the influence of the past information of asset $j$ on its future behaviour. In order to model the influence of the past prices on the future evolution of the time series we include the past returns and the average of the past prices over three different frequencies: day, week and month. In addition, because we work at the daily frequency we want to include information from higher frequency behaviours. When available, we use the open, high, low and close prices as summaries of the day's movement. More specifically, we use three variables constructed from these prices. Let us denote the different versions of the daily price of asset $j$ at time $t$ by $s_{j,t}^{test}$ where $test$ specifies the metric, for example $open$ for the price at the open. Thus, the selected variables representing the intra day moves are: 
\begin{eqnarray*}
r^{low}_t &=& log(s^{low}_t) - log(s^{low}_{t-1}) \, , \\
CH_t &=& \frac{s^{high}_t - s^{close}_t }{s^{high}_t - s^{low}_t} - 0.5 \, , \\
COHL_t &=& \frac{s^{close}_t - s^{open}_t }{s^{high}_t - s^{low}_t} \, .
\end{eqnarray*}
where $r$ represents the log-return. In addition to those variables we will include the leverage effect representing the asymmetric influence of previous positive or negative daily, weekly and monthly return on the future move. Thus the complete vector of endogenous variable is:
\begin{eqnarray*}
e_{j, t} &=& ( 1, as_{j, t}^d, as_{j, t}^w, as_{j, t}^m,\\
&& r_{j, t}^d, r_{j, t}^w, r_{j, t}^m,\\
&& r^{low}_{j,t}, CH_{j,t}, COHL_{j,t},\\
&&  r_{j,t}^{(d)+}, r_{j,t}^{(d)-}, r_{j,t}^{(w)+}, r_{j,t}^{(w)-}, r_{j,t}^{(m)+}, r_{j,t}^{(m)-} ) ^T  \, .
\end{eqnarray*}
Where for each variable the exponent represents the different frequencies: daily, weekly and monthly; $as^f$ represents the averaged price over the frequency $f$; $r^f$ is the past log-return at frequency $f$. Hence the variables of the endogenous vector $e_{j,t}$ represent different economic factors influencing the next price movement.\bigskip

The exogenous variables are composed of a selected group, $sp_t(j)$, of assets which move simultaneously to stock $j$. This part is composed of two steps. The first step proposes candidates for inclusion in $sp_t(j)$, which are then included for a certain time period in the vector of variables. After this time period we look at their signal to noise ratio to decide whether to include it in the group of parents $sp_t(j)$ or not. Once the parents with the highest influence are selected, they are included in the vector of exogenous variables $r_{sp_t(j), t}$. The coefficients $\theta_{j,t}$ of the variables $F_{j,t}$ are then learned sequentially with each new data point following the classic state-space update.

\subsection{Sparsity with the H-SGDLM}
\label{sec:sparsity}
As explained by \cite{dAspremont:2010kh, Cuturi:2013uj, Cuturi:2015wg} sparsity is an important factor to take into account when building mean reverting portfolios. The previously introduced H-SGDLM model constructs a sparse representation of the whole market. Indeed, instead of the classic covariance matrix the cross series relationships are modelled by the exogenous variables; in each individual DLM the exogenous variables inform the mean of the distribution and are sequentially selected and updated for each asset. When combining these individual DLMs to obtain a multivariate distribution, the exogenous coefficients inform both the mean and variance of the multivariate normal through the matrix $\Gamma$. As described previously, the selection of these variables follows a two-step process: first filtering with the Wishart covariance matrix, then testing for its capacity to explain the evolution of the current price given its signal to noise ratio. The number of parents allowed for each DLM is a parameter to be chosen; the smaller this number the greater the sparsity of the resulting multivariate distribution.\bigskip

Thanks to this two steps process, the cross-series relationships are modelled as asymmetric. While some stocks might not be present in any core-group other might be in many. In other words, some assets influence the behaviour of many others in the market while others are mostly followers. These cross-series relationships are represented in the sparse covariance matrix $\Gamma$. Because the number of allowed parents is chosen small, the resulting covariance matrix $\Sigma$ is sparse. And, thanks to its construction the non-zeros elements represent stocks with simultaneous behaviours. Thus, any mean reverting portfolio should be constructed within one of these subsets build from the stocks with the highest non-zero coefficient in the sparse covariance matrix $\Gamma$.

\subsection{MRP reformulated with H-SGDLM data}
We follow the approach of \cite{dAspremont:2010kh, Cuturi:2013uj, Cuturi:2015wg} to minimise the predictability as defined in optimisation problem \ref{eq:minProb}, but instead of using the ratio we will work with the difference:
\begin{eqnarray*}
\text{min}_{x_t} \text{\;\;} x_t^T A^T \Sigma A x_t - x_t^T \Sigma x_t \text{\;\;\; subject to: } && \| x \|_1 = 1 \, , \\
&& \| x \|_0 = k \, .
\end{eqnarray*}
While \cite{dAspremont:2010kh, Cuturi:2013uj, Cuturi:2015wg} modelled the price process by a VAR$(1)$ model (i.e. with stationary autoregressive coefficient $A$ and stationary covariance matrix $\Sigma$), here we use the outputs from the H-SGDLM thereby relaxing this condition of stationarity. By considering the price process $s_t$ to follow a time varying autoregressive process TVAR$(1)$ we can express the portfolio by: $P_t = x_t^T S_t = x_t^T A_t S_{t-1} + x_t^T \epsilon_t$, where the prediction of $S_t$ assuming all the data up to $t-1$ known is denoted by $\hat{S}_t = A_t S_{t-1}$. We recall that the motivation for the \cite{Box:1977iz} predictability ratio was to compare the variance of the prediction to the measured one and see if the prediction dominates the noise, implying that the process is predictable. Thus the ratio of the predicted portfolio variance over the measured one is:
\begin{equation*}
\nu_t = \frac{\hat{\sigma}^2_{t}}{\sigma^2_t} = \frac{\mathbb{E}[ x_t^T A_{t} S_{t-1} S_{t-1}^T A_{t}^T  x_t | D_t]}{\mathbb{E}[ x_t^T S_t S_t^T x_t | D_t]} = \frac{ x_t^T A_{t} \tilde{\Sigma}_{t-1} A_{t}^T  x_t}{x_t^T \tilde{\Sigma}_t x_t}  \, .
\end{equation*}
Where $\tilde{\Sigma}_t$ corresponds to the empirical covariance matrix measured with prices up to $t$. Therefore, this ratio corresponds to the the predicted covariance matrix using the autoregressive coefficient $A_t$ over the measured portfolio variance at $t$. But since our model can actually infer the next covariance matrix $\Sigma_t$ we could use instead:
\begin{equation*}
\nu_t = \frac{x_t^T \Sigma_{t} x_t}{x_t^T \tilde{\Sigma}_t x_t} \, .
\end{equation*}
The H-SGDLM model gives us the autoregressive coefficient $A_t$, and sparse predicted covariance matrix $\Sigma_t$. Therefore the predictability ratio could be computed in different ways using either one of these outputs. Since for the optimisation problem we work with the difference instead of the ratio, possible utility functions include:
\begin{eqnarray*}
U_1(x) &=& x^T \left( A_{t} \tilde{\Sigma}_{t-1} A_{t}^T -  \tilde{\Sigma}_t \right) x \, , \\
U_2(x) &=& x^T \left( \Sigma_{t} -  \tilde{\Sigma}_t \right) x \, , \\
U_3(x) &=& x^T \left( A_{t} \tilde{\Sigma}_{t-1} A_{t}^T -  \Sigma_t \right) x \, .
\end{eqnarray*}
Let us detail the different options. $U_1$ uses the measured covariance matrices at $t-1$ and $t$ and computed autoregressive coefficient $A_t$. $U_2$ compares the sparse covariance matrix at $t$ inferred with data up to $t-1$ to the measured covariance matrix using prices up to $t$. $U_3$ compares a prediction of the variance using the covariance matrix measured with data up to $t-1$ and the computed autoregressive coefficients to the predicted covariance matrix $\Sigma_t$. Now since we assumed all the data up to $t$ is known for computing the predictability ratio at $t$, we could do the same computation as $U_3$ but using the data available at $t$, i.e. compare the $t+1$ predictions. However, this option (as with $U_3$) does not respect the predictability as defined by \cite{Box:1977iz} since it would compare two predictions instead of a predicted versus a measured value; for this reason we work with $U_1$ and $U_2$. With this set-up we can either use the predicted covariance matrix or the predicted autoregressive coefficient. Let us denote the minimisation of the covariance differences as problem $PC$ and the other one as problem $PA$. Further, let us denote the utility function by $x^T \left( D_t - \tilde{\Sigma}_t \right) x$, where $D_t$ is the predicted covariance matrix. Then, for $PC$ we have $D_t= \Sigma_t $, while for $PA$ it is $D_t=A_{t} \tilde{\Sigma}_{t-1} A_{t}^T$.

\section{Solving the minimisation problem with CCD}
\label{sec:CCD}
\subsection{The CCD algorithm for non-convex function}
Coordinate descent algorithms work by optimising one direction at a time instead of all of them simultaneously as classical descent algorithms do. The main drawback for this type of algorithm is their requirement for the utility function to be strictly convex and differentiable. \cite{Tseng:2001bd} proved that the convergence can be extended to a specific type of non-convex and non-differentiable function. In particular they considered a non-convex objective function that can be decomposed into a quasi-convex and differentiable function $f_0$ and a non-differentiable but additively separable one $f_k$:
\begin{equation*}
f(w) = f_0(w) + \sum_{k=1}^{m} f_k(w_k) \, .
\end{equation*}
If a function can be written in such a way, then a block coordinate descent algorithm applied on this function will converge to the global minimum of $f$ with respect to $w$. More specifically, the convergence is guaranteed according to theorem $5.1$ of \cite{Tseng:2001bd} if the functions respect the following conditions: $f_0$ must be continuous, $f$ is quasi-convex and hemivariate in each coordinate block and the functions $f_k$ are left side continuous $\forall k$. In addition, \cite{Tseng:2001bd} made assumptions on the domain of $f_0$, for which we refer the reader to the original paper.\bigskip

The coordinate descent algorithm works by iteratively updating the weight vector $w^r$. At each step it updates the value of the weights $w^r_i$ for all $i\in [1, N]$ by minimising $f$ assuming all the other weights $w^r_{j \neq i}$ fixed: $w^r_i = min_{w^r_i} f(w^r)$. Then, it minimises the next weights using the updated values of the other weights. Once the entire vector $w^r$ has been updated the iteration starts again until convergence: $\|w^r - w^{r-1} \| \leq \epsilon$.\bigskip

Subsequently, \cite{Anonymous:2014di} used this algorithm to solve the minimum variance portfolio optimisation with elastic net constraints since it naturally fits in the formulation of \cite{Tseng:2001bd}. The Lagrangian of this problem is:
\begin{equation*}
L\left(W, \gamma, \Sigma, \beta, \alpha \right) = W^T \Sigma W +  \beta \left( \alpha \|W \|_{l_1} + \left( 1 - \alpha \right)  \|W \|_{l_2}^2 \right) - \gamma \left(W^T1 -1 \right) ,
\end{equation*}
where $\beta$ is the penalty parameter, $\alpha$ the weight balancing the importance of the $L_1$ and $L_2$ terms, $\gamma$ the Lagrange multiplier and $W$ the weights vector. Fixing $\gamma$, we can solve the KKT conditions with respect to the weight $w_i$ to obtain:
\begin{eqnarray*}
w_i &=& \frac{ST\left( \gamma - z_i, \beta \alpha \right)}{2 \left(\sigma_i^2 + \beta \left(1-\alpha \right) \right)} \, , \\
z_i &=& 2 \sum_{j\neq i}^{N} w_j \sigma_{i,j} \, ,
\end{eqnarray*}
where $\sigma_{i,j}$ corresponds to the covariance between weights $i$ and $j$, and $ST(x, y) = sign(x) max(|x| - y, 0 )$ is the soft-threshold function. The CCD algorithm recursively updates each weight $w_i$ with this equation until the vector $w$ converges. The following section will detail the application of the CCD algorithm to solve the sparse mean reverting portfolio optimisation problem. 

\subsection{The CCD algorithm applied to sparse mean reverting portfolios}
Instead of the cadinality constraint used by \cite{dAspremont:2010kh, Cuturi:2013uj, Cuturi:2015wg} to fix the number of non-zeros weights to a pre-selected value, we use the graphical structure learned by the H-SGDLM model to select a group of assets of size $k$. While in this paper we use the predicted covariance matrix $D_t$ obtained from the H-SGDLM model, the CCD algorithm we propose in this section could be applied using any other covariance matrix prediction, however we note that without the H-SGDLM sparsity selection, an $L_1$ regularisation coefficient would be needed to impose sparsity. Since we work with the difference instead of the ratio we add a $L_2$ regularisation factor to the utility function. The minimisation problem becomes:
\begin{equation*}
\label{eq:CCDminProb}
\text{min}_{x_t} \text{\;\;} x_t^T \left( D_t - \tilde{\Sigma}_t \right) x_t + \beta \| x_t \|_2^2  \, , \text{\;\;\; subject to: \;\;} \| x_t \|_1 = 1 \, ,
\end{equation*}
where $\beta$ is the weight of the $L_2$ norm. With this formulation only the normalisation constraint remains. We use the $L_1$ norm of the weight as the normalisation constraint because we want to take into account the full exposure of the portfolio. Hence, the normalisation takes into account the short positions in the total weight. In addition to its smoothness property, the $L_2$ regularisation term here guarantees the positive semi-definiteness of the problem. Indeed, although $D_t$ and $\Sigma_t$ are positive semi-definite covariance matrices, their difference is not guaranteed to be. We circumvent this issue with the addition of an $L_2$ weight, which acts as a matrix regularisation coefficient to keep the difference within the positive semi-definite domain.\bigskip

This approach is not perfect however, since the $L_2$ constraint adds a bias to the solution. Therefore we want to select the smallest possible coefficient $\beta$ that guarantees the positive semi-definiteness of the matrix $M=D_t - \tilde{\Sigma}_t + \beta I$, where $I$ corresponds to the identity matrix. We select a value $\beta=\beta_0$ and at each time $t$ compute the optimal solution, then we check for the positivity of the resulting utility function. In case of negativity, we increase the value of $\beta$ by $10^{-5}$ until the function becomes positive, but only for a maximum of $10$ steps. If the maximum number of iterations is reached we increase the step-size to $10^{-1}$ and start again. This incremental approach is computationally feasible due to the speed of the CCD algorithm, although we have to take care with choice of $\beta_0$; if it is chosen too small it will end up in the loop step too often, while if chosen too big it adds an unnecessary bias. In addition, if the value of the MRP part becomes negligible compared to the value of the $L_2$ constraint then the optimisation algorithm will just minimise the $L_2$ norm of the weight and thus produce the equal weighted portfolio. We detail in section \ref{sec:results} a pragmatic selection of values for the problems considered.\bigskip

For the sake of clarity, since we consider all the data up to $t$ known and assume all the variables in the optimisation problem are taken at time $t$, we drop the subscript $t$ in the following expressions. The Lagrangian of our minimisation problem (\ref{eq:CCDminProb}) follows as:
\begin{equation}
\label{eq:lagrangian}
\mathcal{L} (x, \gamma) = x^T \left( D - \tilde{\Sigma} + \beta I \right) x  +\gamma \left(  \| x \|_1  - 1 \right) \, ,
\end{equation}
This function satisfies the convergence requirements defined by \cite{Tseng:2001bd} since we can write the problem as the combination of a differentiable, quasi-convex term and a non-differentiable but additively separable one. Thus, we can follow a coordinate descent to find the global optimum. Without loss of generality we will fix the Lagrange coefficient to zero and renormalise the weight vector at the end of each loop. Let us denote the elements of matrix $D$ by $d_{i,j}$ and of $\Sigma$ by $\sigma_{i,j}$. Then, the line $i$ of the matrix $\left( D - \tilde{\Sigma} + \beta I \right) x $ can be written $\left( D - \tilde{\Sigma} + \beta I  \right) x |_{i} =  \left( d^2_{i} - \sigma^2_{i} + \beta \right) x_i + \Sigma_{j \neq i} \left( d_{i,j} - \sigma_{i,j} \right) x_j $. Where $\sigma^2_{i} = \sigma_{i, i}$ and $d^2_{i} = d_{i,i}$ are the diagonal elements of $D$ and $\Sigma$ respectively. Thus, the gradient of the Lagrangian in Equation \ref{eq:lagrangian} with respect to $x_i$ is:
\begin{equation*}
\nabla_{x_i} \mathcal{L} (x, \gamma) = 2  \left( d^2_{i} - \sigma^2_{i} + \beta \right) x_i  + 2 \Sigma_{j \neq i} \left( d_{i,j} - \sigma_{i,j} \right) x_j \, ,
\end{equation*}
Therefore, for the gradient to be zero each weight $x_i$ must follow:
\begin{equation}
\label{eq:ccdUpdate}
x_i = -\frac{2\Sigma_{j \neq i} \left( d_{i,j} - \sigma_{i,j} \right) x_j }{2 \left(  d^2_{i} - \sigma^2_{i}  +\beta \right)} \, .
\end{equation}
The coordinate descent algorithm iteratively updates the weights with equation \ref{eq:ccdUpdate} until convergence. While we did not include an $L_1$ norm on the weights because we enforce sparsity directly using the results from the H-SGDLM model, it is straightforward to include one. Indeed, with an $L_1$ regularisation term and coefficient $\lambda_1$, the numerator of Equation \ref{eq:ccdUpdate} simply becomes ST $\left( 2\Sigma_{j \neq i} \left( d_{i,j} - \sigma_{i,j} \right) x_j , \lambda_1 \right)$, where $ST(., .)$ is the soft-thresholding function.

\section{Trading a Mean Reverting Portfolio}
\label{sec:trading}
\subsection{Jurek and Yang trading strategy}
In the papers \cite{Cuturi:2013uj, Cuturi:2015wg} used the \cite{Jurek:2007ju} strategy to trade the mean reversion of the obtained MRP portfolios. This strategy considers the portfolio to follow a stationary VAR$(1)$ process $P_t = \rho P_{t-1} + \sigma \epsilon_t$, with $|\rho| < 1$  and mean $\mu$, where the parameters were obtained using a least squares estimate. Let us denote by $w_t$ the investment weight of portfolio $P_t$, at $t$, which the authors define by $w_t = (\mu - p_t) \rho / \sigma^2 $. In their paper from 2013 the authors considered trading swaps and concluded that the gains are too small to create a profitable strategy in practice, since the transaction costs cancelled all arbitrage opportunities. On the other hand, in the 2015 paper they built baskets of options to trade implied volatility and with added constraints they obtained performances robust to costs.

\subsection{Ornstein-Ulhenbeck based trading strategy}
In their papers, \cite{Fogarasi:eh, Sipos:2013ke} assumed the portfolio to follow an Ornstein-Ulhenbeck process $P_t \sim \mathcal{N} \left( \mu, \sqrt{\frac{\sigma^2}{2 \rho}} \right)$ and considered three scenarios. Either the portfolio is in a stationary state and no investment is made, or it is outside a threshold and a position is entered. The authors define a threshold $\epsilon$ such that:
\begin{itemize}
\item H1: $P\left( p_t < \mu - \alpha \right) = \epsilon /2$, the portfolio's value is bellow its long term mean and they buy it.
\item H2: $P\left( p_t \in [ \mu - \alpha, \mu + \alpha ]\right) = \epsilon $, the portfolio is in its stationary state and thus they stay neutral.
\item H3: $P\left( p_t > \mu + \alpha \right) = \epsilon /2$, the portfolio's value is above its long term mean and they sell it.
\end{itemize}
Once a position is entered, they keep it until the portfolio comes back to its stationary state. In this strategy based on O-U thresholds the maximum amount is always invested, in contrast to the original strategy of \cite{Jurek:2007ju} where the portfolio's weight can take any real number.

\subsection{Mixed mean reversion trading strategy}
We could also use a combination of those two ideas. Instead of always investing an amount corresponding to the distance from the mean (as in \cite{Jurek:2007ju}), we could instead adopt this approach only when the move is larger than a threshold, resulting in the strategy:
 \begin{itemize}
\item H1: $W_t = \frac{\rho}{\sigma^2} \left((\mu - \alpha) - P_t \right)$,
\item H2: $W_t = 0$,
\item H3: $W_t = \frac{\rho}{\sigma2} \left( (\mu + \alpha) - P_t\right)$,
\end{itemize}
For the rest of this paper we will refer to this strategy as the mixed $JY-OU$ strategy.

\section{Results}
\label{sec:results}
\subsection{US stocks from the S\&P$500$}
In order to test our algorithm on stocks, we selected the $371$ stocks from the S\&P$500$ with data available since $2001$. This dataset consists of the OHLCV data for each stock at the end of each trading day. However, instead of using daily time points we worked on weekly time points, i.e. we selected one data every $5$ time points. Hence, the one-step ahead inference on which the H-SGDLM learns corresponds to a one week ahead forecast. Since the H-SGDLM takes some time to converge we study the results between $2003$ and $2018$. The vector of variables includes the OHLC data as described in section \ref{sec:HSGDLM}. The size of the core parent group was fixed to $10$ with an update every $10$ time steps and due to the memory limit of the GPU card the number of Monte-Carlo samples was fixed to $500$.\bigskip

\begin{figure}[ht]
\centering
\includegraphics[width = \figurewidth, height = \figureheight]{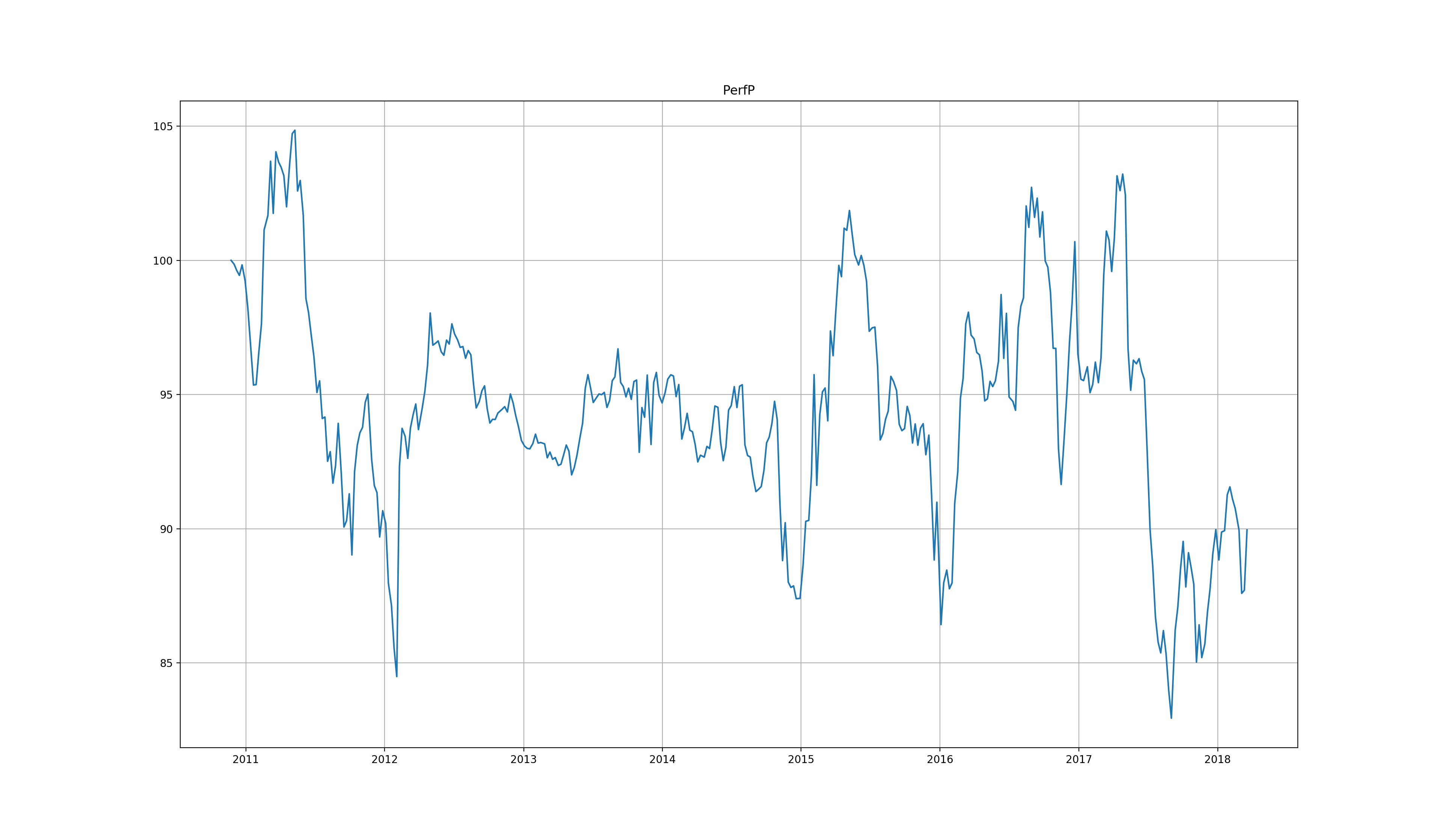}
\vspace{-20pt}
\caption{\label{fig:USport11} This figure shows the performance of the MRP portfolio computed for $371$ stocks from the S\&P$500$. On this figure we can observe the mean reversion of the obtained portfolio.}
\end{figure}

\begin{figure}[ht]
\centering
\includegraphics[width = \figurewidth, height = \figureheight]{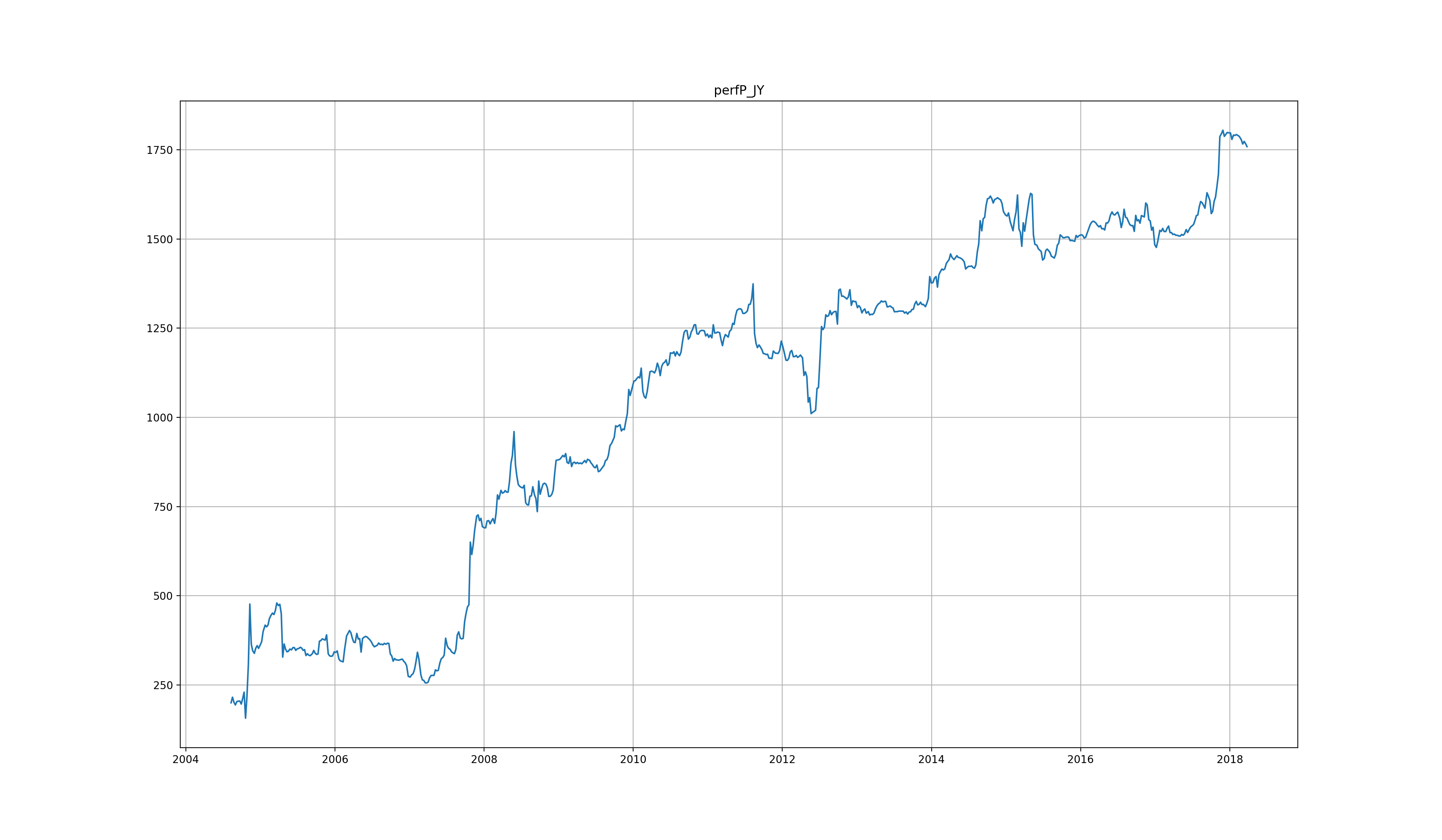}
\vspace{-20pt}
\caption{\label{fig:USportJY} This plot shows the performance of the \cite{Jurek:2007ju} trading strategy applied on the MRP portfolio computed on an evolving set of $50$ stocks from the S\&P$500$.}
\end{figure}

\begin{figure}[ht]
\centering
\includegraphics[width = \figurewidth, height = \figureheight]{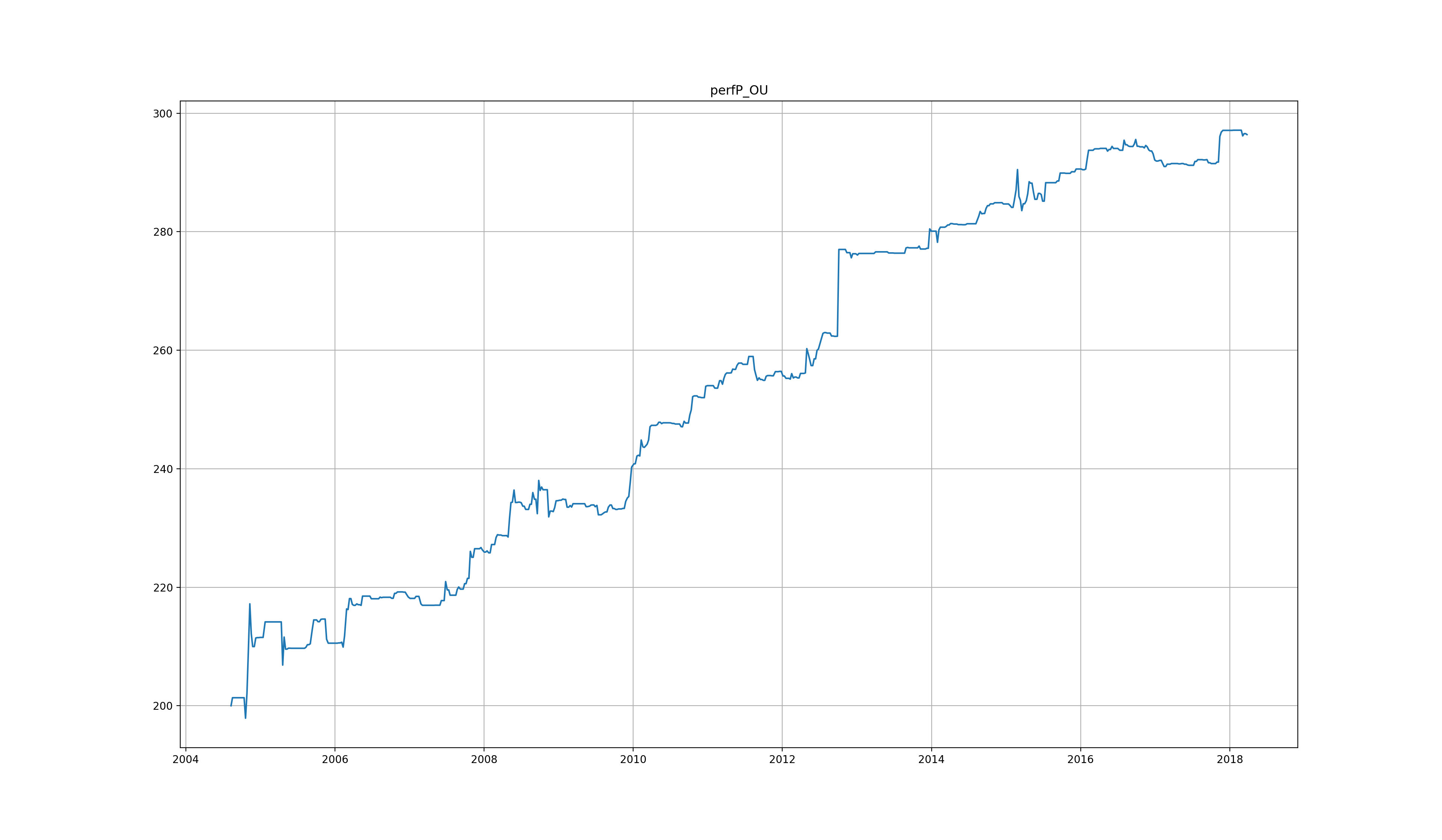}
\vspace{-20pt}
\caption{\label{fig:USportOU} This plot shows the performance of the O-U based trading strategy applied on the MRP portfolio computed on an evolving set of $50$ stocks from the S\&P$500$.}
\end{figure}

For the parametrisation step we selected half the dataset, from $2011$ to $2018$. We used the second half of the dataset to avoid fitting the parameters over the $2008$ crisis, as we wanted our out-of-sample to test the robustness of the parameters over this crisis period. The MRP algorithm has two parameters, $\beta$ the $L_2$ weight and $\lambda$ the coefficient of the exponential weighting we use to compute the empirical covariance matrix $\tilde{\Sigma}$. Since we use two different optimisation problems, $PA$ and $PC$, we allow the parameters to be different for each one. To find the acceptable range of values for those parameters we compared the computed MRP portfolios to the equally weighted one. Once this range was found, we selected the value that created the most mean-reverting portfolio over this time window. As an example figure \ref{fig:USport11} shows the resulting portfolio obtained over the whole dataset with $\beta = 10^{-5}$ and $\lambda=0.9$. Since we want a sparse portfolio, we fixed the size of the portfolio to $50$ different assets following the methodology described in \ref{sec:sparsity} that uses the matrix of parents' coefficients $\Gamma$. We then selected $\beta=\{10^{-3}, 10^{-4}, 10^{-5}, 10^{6} \}$, $\lambda_{PA}=0.98$ and $\lambda_{PC}=0.85$.\bigskip
For the trading part we have to select the length, $T_{tr}$ of the time window on which the different trading parameters will be computed at each time $t$. We looked at the performances of the backtests and chose $T_{tr}=50$. We then compared the different trading strategies described in \ref{sec:trading}. For each strategy we averaged the results obtained for the different $\beta$ values and the different $PA$ and $PC$ problems. The figures \ref{fig:USportJY} and \ref{fig:USportOU} show the results for the \cite{Jurek:2007ju} and O-U based strategies respectively. Interestingly on this dataset the mixed $JY-OU$ one had worse performance then the other two. Even though the backtests do not include any costs it is interesting to observe the steady performance even during the $2008$ crisis. 

\subsection{FX Futures}
In this section we test our algorithm on exchange rates data with a dataset consisting of $22$ different FX-futures between $2006$ and $2018$. For this environment, the H-SGDLM used the average of the past $40$ daily-, weekly-, monthly-log-returns. This average used an exponential weighting with a coefficient of $0.98$. The size of the parent group was fixed at $5$ and the updates happened every $10$ time steps. Just like in the previous example, we used a weekly time point dataset, such that the parent selection corresponded to an update every $50$ days. Furthermore, we used $2000$ Monte-Carlo samples.\bigskip

\begin{figure}[ht]
\centering
\includegraphics[width = \figurewidth, height = \figureheight]{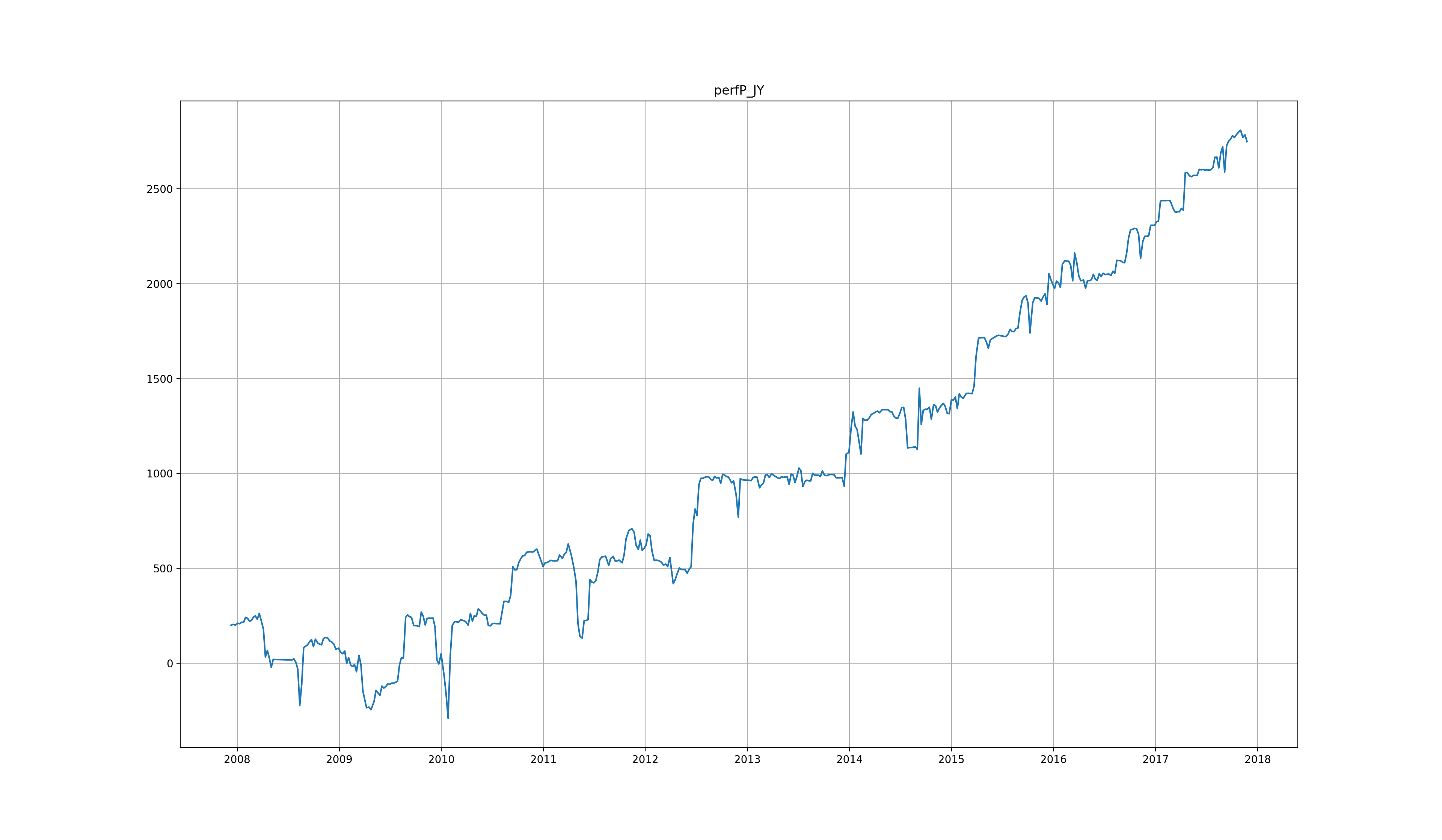}
\vspace{-20pt}
\caption{\label{fig:FXportJY}This plot shows the performance of the \cite{Jurek:2007ju} trading strategy applied on the MRP portfolio computed on an evolving set of $10$ FX-futures.}
\end{figure}

\begin{figure}[ht]
\centering
\includegraphics[width = \figurewidth, height = \figureheight]{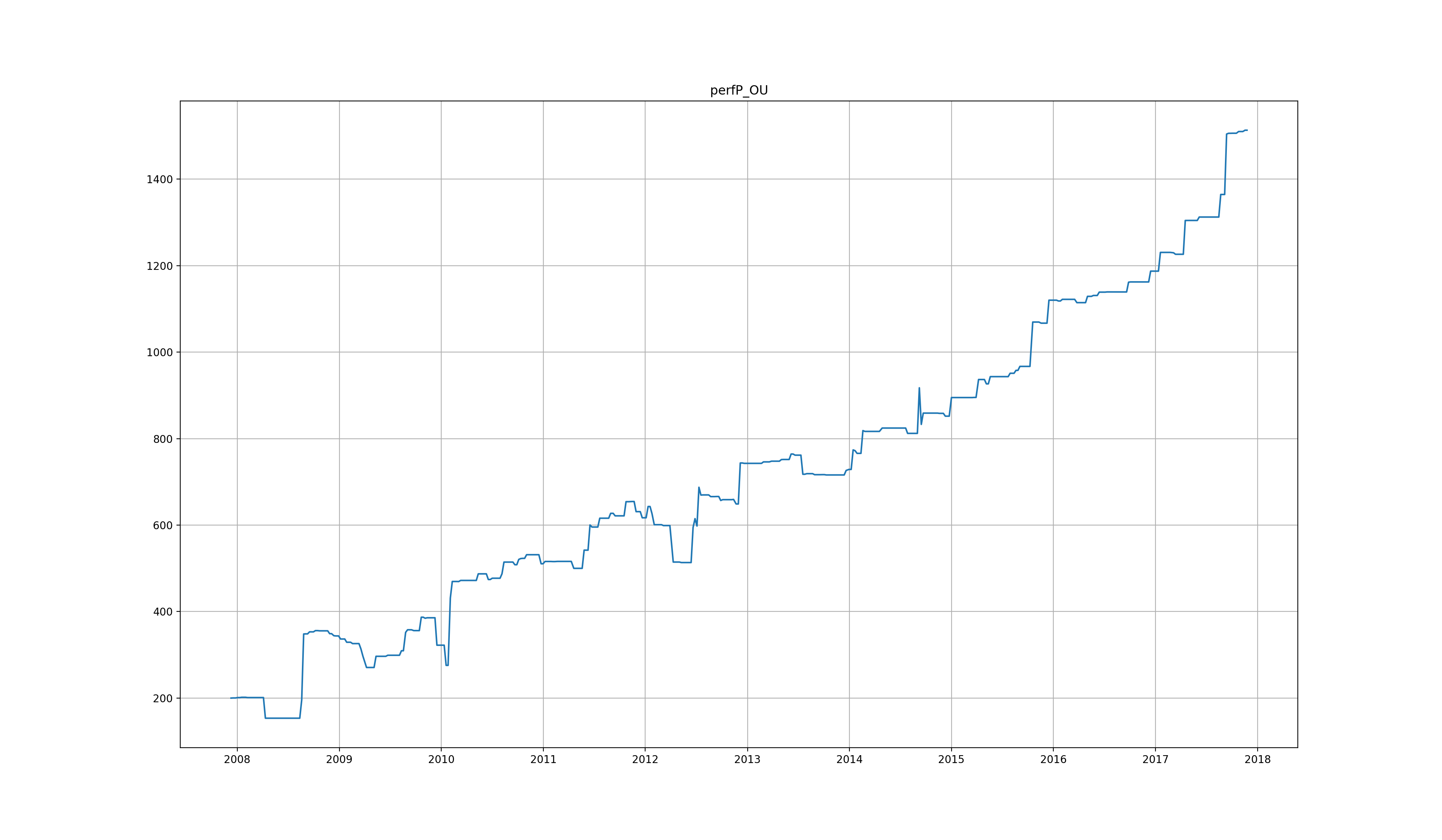}
\vspace{-20pt}
\caption{\label{fig:FXportJYOU}This plot shows the performance of the mixed $JY-OU$ trading strategy applied on the MRP portfolio computed on an evolving set of $10$ FX-futures.}
\end{figure}

We followed the same methodology to find the parameters as the one described for S\&P$500$ data. Hence, the parametrisation step used half the dataset, from $2012$ to $2018$ and the number of asset of the portfolio, or sparsity constraint, is fixed to $10$. The exponential weighting coefficient is chosen to be the same for both problems $PA$ and $PC$, i.e. $\lambda_{PA}=\lambda_{PC}=0.8$. But, the $L_2$ coefficients are different: $\beta_{PA}=\{10^{-1}, 10^{-2}, 10^{-3} \}$ and $\beta_{PC}=\{10^{-2}, 10^{-4}, 10^{-5} \}$. For the trading strategies the time window was fixed to $T_{tr}=10$ time points equal to $50$ trading days. Figure \ref{fig:FXportJY} shows the backtest performance using the \cite{Jurek:2007ju} trading strategy while figure \ref{fig:FXportJYOU} corresponds to the backtest of the mixed $JY-OU$ strategy. As with the previous backtest on stocks it is interesting to observe the algorithm's steady performance. In this case however, contrary to the backtest on stocks, the mixed $JY-OU$ strategy performed slightly better than the O-U based one, hence we only show the performance of the mixed one.

\subsection{ETF Futures}
Finally we considered the dataset of futures on ETFs, in particular $75$ different futures with OHLCV data from $2008$ to $2018$. Hence, the H-SGLDM learns on log-prices with the variables described in section \ref{sec:HSGDLM}. As for the previous datasets we used weekly time points instead of daily ones. The size of the core parent group is fixed to $10$ with an update every $10$ time steps and we computed $2000$ Monte-Carlo samples.\bigskip

\begin{figure}[ht]
\centering
\includegraphics[width = \figurewidth, height = \figureheight]{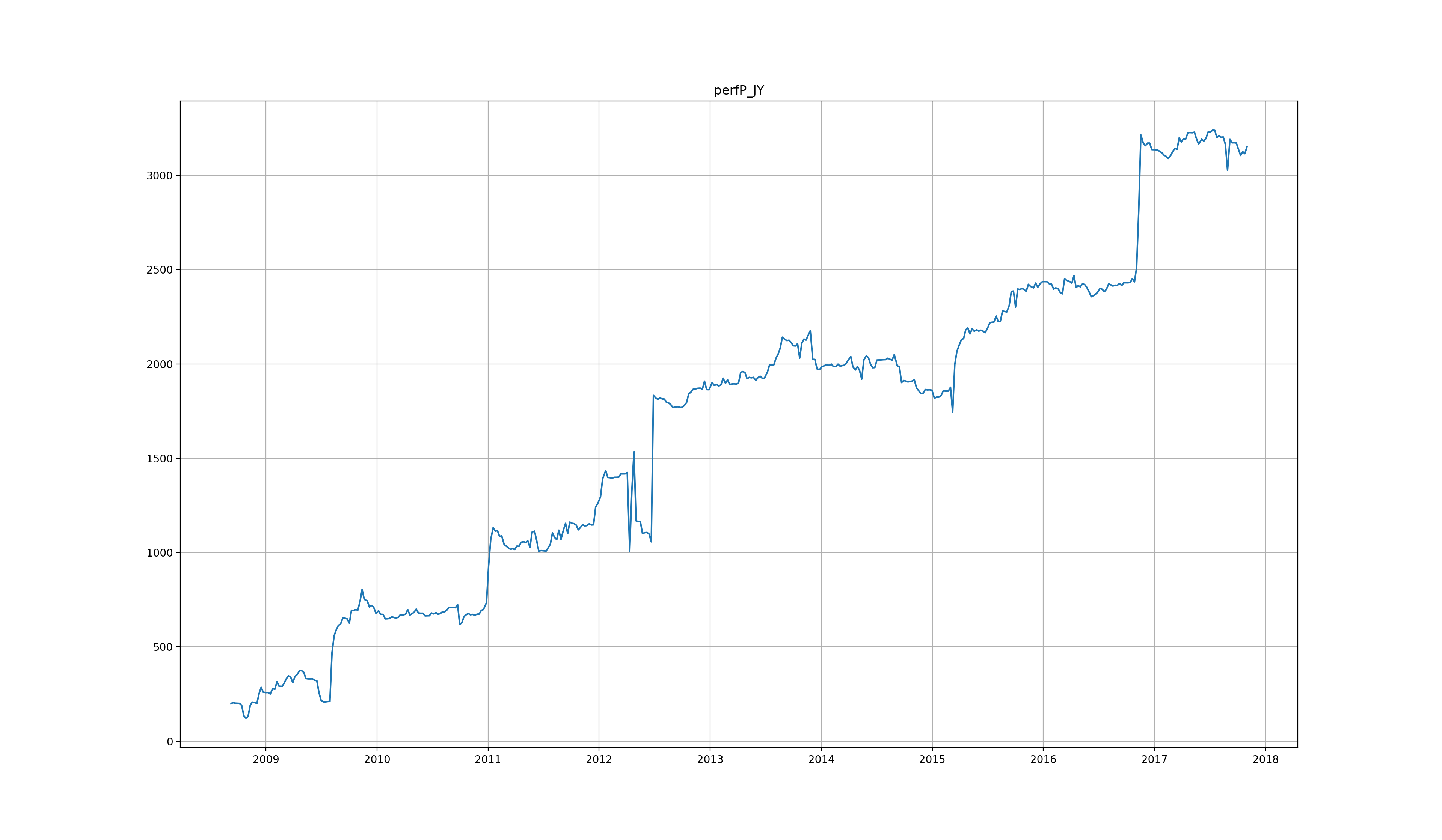}
\vspace{-20pt}
\caption{\label{fig:ETFportJY}This plot shows the performance of the \cite{Jurek:2007ju} trading strategy applied on the MRP portfolio computed on an evolving set of $20$ ETF-futures.}
\end{figure}

\begin{figure}[ht]
\centering
\includegraphics[width = \figurewidth, height = \figureheight]{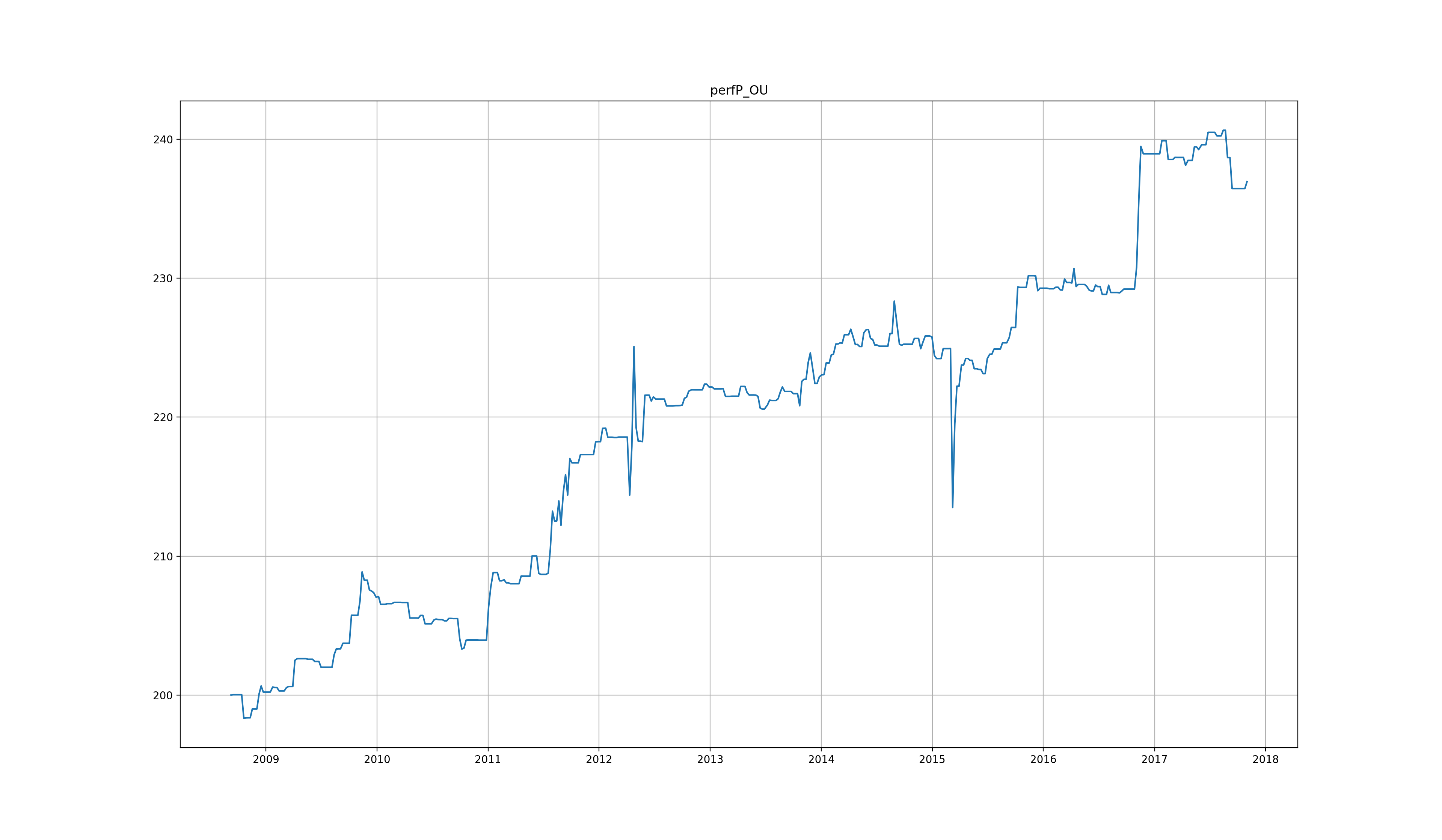}
\vspace{-20pt}
\caption{\label{fig:ETFportJYOU}This plot shows the performance of the O-U based trading strategy applied on the MRP portfolio computed on an evolving set of $20$ ETF-futures.}
\end{figure}

We followed the same methodology as before to find the MRP parameters. The sparsity variable was fixed to $20$ futures, while the time window used for parametrisation was $2013$ to $2018$. We selected exponential weighting coefficients of $\lambda_{PA}=0.85$ and $\lambda_{PC}=0.98$, and $L_2$ coefficients: $\beta_{PA}=\{10^{-5}, 10^{-6} \}$ and $\beta_{PC}=\{10^{-3}, 10^{-4} \}$. As with FX-futures the time window for the trading strategies was fixed to $T_{tr}=10$ time points. As for the dataset on stocks, the mixed $JY-OU$ strategy performed worse than the other two. Figure \ref{fig:ETFportJY} shows the backtest of the \cite{Jurek:2007ju} trading strategy and figure \ref{fig:ETFportJYOU} corresponds to the performance of the O-U based approach. While these backtests are more volatile than the those on FX-futures and US-stocks they still have a relatively steady growth highlighting the robustness of this portfolio construction.

\section{Conclusion}
\label{sec:Conclusion}
In this paper we presented a new approach for computing a sparse mean reverting portfolio within a large environment. The goal was to find an efficient procedure that is able to find a sparse subset of series from a large environment of time series, for which we could compute a linear combination exhibiting the mean-reverting property.  Furthermore, the solution had to work on the more complicated problem of subsets larger than two assets.  Given such a mean-reverting portfolio the performance of our solution was numerically evaluated by performing backtests in different investment environments.\bigskip

The solution we proposed in this paper leveraged the data obtained by the newly introduced H-SGDLM model. Indeed, we were able to reformulate the MRP problem as a quasi-convex optimisation problem with a normalisation constraint.  In addition, the sparsity constraint was moved away from the minimisation formulation and instead we used the matrix of core parents from the H-SGDLM to select the subset of time series on which to compute the MRP.  Using different variables from the H-SGDLM model we obtained two different formulations of the problem, $PA$ and $PC$, using the matrix of the regression coefficients and the predicted sparse covariance matrix, respectively. This new formulation is perfectly suited for a cyclical coordinate descent (CCD) algorithm. Hence, once the data from the H-SGDLM is available the CCD optimisation step finds the global solution of the MRP minimisation problem.\bigskip

As a result of this formulation it is possible to compute the optimal MRP at each time $t$ even on environments involving hundreds of time series. We then evaluated different trading strategies to take advantage of this mean-reversion. We performed backtests on three different datasets: stocks from the S\&P$500$, FX-futures and ETF-futures. All the resulting backtests exhibited steady performances through time even on the out-of-sample part of the backtest that included the financial crisis, suggesting a certain robustness of our proposed solution.

\bibliographystyle{rQUF}
\bibliography{bibTexLib_MRP}

\appendix

\end{document}